\documentclass[twocolumn,printnumbers,amsmath,amssymb,showpacs,prl]{revtex4}
\usepackage{graphicx}% Include figure files
\usepackage{color}

\begin{document}
\title{Disordered Solids Without Well-Defined Transverse Phonons: The Nature of Hard Sphere Glasses}

\date{\today}

\author{Xipeng Wang}
\author{Wen Zheng}
\author{Lijin Wang}
\author{Ning Xu$^*$}

\affiliation{CAS Key Laboratory of Soft Matter Chemistry, Hefei National Laboratory for Physical Sciences at the Microscale, and Department of Physics, University of Science and Technology of China, Hefei 230026, People's Republic of China.}

\begin{abstract}
We probe the Ioffe-Regel limits of glasses with repulsions near the zero-temperature jamming transition by measuring the dynamical structure factors.  At zero temperature, the transverse Ioffe-Regel frequency vanishes at the jamming transition with a diverging length, but the longitudinal one does not, which excludes the existence of a diverging length associated with the longitudinal excitations.  At low temperatures, the transverse and longitudinal Ioffe-Regel frequencies approach zero at the jamming-like transition and glass transition, respectively.  As a consequence, glasses between the glass transition and jamming-like transition, which are hard sphere glasses in the low temperature limit, can only carry well-defined longitudinal phonons and have an opposite pressure dependence of the ratio of the shear modulus to the bulk modulus from glasses beyond the jamming-like transition.
\end{abstract}


\maketitle

Upon compression, colloidal systems undergo the glass transition when the relaxation time (or the viscosity) exceeds the measurable value \cite{debenedetti,berthier1,ediger}.  In the absence of the thermal energy, the compression leads to the jamming transition of packings of repulsive particles with the sudden formation of rigidity and static force networks \cite{liu,ohern}.  Although the initial jamming phase diagram \cite{liu} proposes that the glass transition of purely repulsive systems collapses with the jamming transition in the zero temperature ($T=0$) limit, or equivalently in the hard sphere limit \cite{xu1}, recent studies have evidenced that the $T=0$ or hard sphere glass transition happens at a packing fraction $\phi_{g0}$ lower than the critical value $\phi_{j0}$ of the $T=0$ jamming transition at the so-called point $J$ \cite{zhang,wang,krzakala,berthier2,parisi,ikeda}.

The departure of $\phi_{g0}$ from $\phi_{j0}$ leaves $\phi\in (\phi_{g0},\phi_{j0})$ a special region.  At $T=0$, the systems in this region are unjammed and unable to sustain the shear or compression.  However, they are by definition glasses when being thermally excited, because particles are unable to diffuse freely.  While glasses are believed to be mechanically rigid solids, it seems perplexing that the systems in $(\phi_{g0},\phi_{j0})$ are rigid in the glass perspective but not in the $T=0$ jamming perspective, which makes $T=0$ singular.

For soft and repulsive spheres, recent simulations have indicated that the glass transition temperature $T_g$ is proportional to the pressure $p$ and vanishes at $\phi_{g0}$ \cite{zhang,wang,xu1,berthier2}.  Both colloidal experiments and simulations have also shown that inside the glass regime, at fixed temperature, there exists a crossover pressure $p_j$ at which the first peak of the pair distribution function reaches the maximum height $g_1^{{\rm max}}$ \cite{zhang,wang,berthier3,cheng,yunker}, reminiscing the structural signature of the $T=0$ jamming transition \cite{silbert1}. At such a crossover, the pressure dependence of the temperature $T_j(p)$ is different from $T_g(p)$.  $T_j(p)$ is lower than $T_g(p)$ and vanishes at $\phi_{j0}$ \cite{zhang,wang}. In other words, when compressed at a fixed low temperature, the system undergoes the glass transition first and then the emergence of $g_1^{{\rm max}}$. The whole picture is reproduced here as well in Fig.~\ref{fig:fig3}.

A recent study has demonstrated further that at $p>p_j$ and at a fixed low temperature some typical scaling relations observed in the $T=0$ jammed solids are recovered, while they break down at $p<p_j$, so it has been suggested that the emergence of $g_1^{\rm max}$ at $p_j$ signifies the jamming-like transition \cite{wang}. Solid properties of glasses on both sides of the jamming-like transition must be distinct.  For instance, the thermal energy softens jammed solids, but it hardens the systems that are unjammed at $T=0$ by increasing the yield stress with the temperature \cite{ikeda}.  We will show that the underlying mechanism of such distinctions challenges our understanding of glasses with repulsions, especially the nature of hard sphere glasses.

By measuring the dynamical structure factors of soft glasses with repulsions, we obtain both the transverse and longitudinal Ioffe-Regel frequencies, i.e., the critical frequencies at which the phonon wavelength is comparable to the mean free path \cite{ioffe}.  At $T=0$, the transverse Ioffe-Regel frequency $\omega_{IR}^T$ vanishes at point $J$, while the longitudinal one $\omega_{IR}^L$ remains nonzero.  Our analysis of the low-temperature glasses suggests that $\omega_{IR}^T$ and $\omega_{IR}^L$ vanish with a diverging length at the jamming-like transition and glass transition, respectively.  Therefore, glasses below the jamming-like transition do not have well-defined transverse phonons, which should be the intrinsic cause of their significantly different material properties from glasses above the jamming-like transition.

Our systems are $L\times L$ boxes in two dimensions with periodic boundary conditions.  To avoid crystallization, we put in the box $N$ bidisperse disks with an equal mass $m$ and a diameter ratio $1.4$.  If not specified, the results shown are for $N=4096$ systems.  Particles $i$ and $j$ interact via purely repulsive potential
\begin{equation}
U_{ij} = \frac{\epsilon}{\alpha}\left( 1 - \frac{r_{ij}}{\sigma_{ij}}\right)^{\alpha} \Theta\left( 1 - \frac{r_{ij}}{\sigma_{ij}}\right),
\end{equation}
where $r_{ij}$ is their separation, $\sigma_{ij}$ is the sum of their radii, and $\Theta(x)$ is the Heaviside function.  We show here the results of Hertzian repulsion ($\alpha=5/2$), which has been verified as a good approximation to the interaction of PNIPAM colloids \cite{zhang}.  In our simulations, the length, energy, and mass are in units of small particle diameter $\sigma$, characteristic energy of the interaction $\epsilon$, and particle mass $m$.  The time is in units of $\sqrt{m\sigma^2/\epsilon}$.  The temperature is in units of $\epsilon/k_B$ with $k_B$ the Boltzmann constant.

We generate jammed solids at $T=0$ by finding the local potential energy minima using the fast inertial relaxation engine minimization algorithm \cite{fire}.  The normal modes of vibration are obtained by diagonalizing the Hessian matrix using ARPACK \cite{arpack}.  The dynamical structure factors at $T=0$ are measured from the modes \cite{shintani}:
\begin{equation}
S_{\lambda}(k, \omega) = \frac{k^2}{m \omega^2}\sum_n F_{n,\lambda}(k) \delta (\omega - \omega_n),\label{s1}
\end{equation}
where the sum is over all modes, $\lambda$ denotes $T$ (transverse) or $L$ (longitudinal), and
\begin{eqnarray}
F_{n,L} &=& |\sum_j (\vec{e}_{n,j}\cdot \hat{k}) {\rm exp}({\rm i}\vec{k}\cdot \vec{r}_j)|^2,\label{fl}\\
F_{n,T} &=& |\sum_j (\vec{e}_{n,j}\times \hat{k}) {\rm exp} ({\rm i}\vec{k}\cdot \vec{r}_j)|^2, \label{ft}
\end{eqnarray}
where the sums are over all particles, $\vec{e}_{n,j}$ is the polarization vector of particle $j$ in mode $n$, $\vec{r}_j$ is the location of particle $j$, and $\hat{k}= \vec{k} / k$ with $\vec{k}$ satisfying the periodic boundary conditions.

For thermal systems, we perform molecular dynamics simulations at constant temperature and pressure.  The structure and dynamics are evaluated from the pair distribution function $g(r)=(4L^2/N^2)\langle \sum_i\sum_{j\ne i}\delta(r-r_{ij})\rangle$ and intermediate scattering function $F_s(q,t)=(2/N)\langle\sum_{j} {\rm exp}({\rm i}\vec{q}\cdot[\vec{r}_{j}(t)-\vec{r}_{j}(0)])\rangle$ for large particles, where the sums are over all large particles, $\vec{q}$ is chosen in the $x$ direction with the amplitude approximately equal to the value at the first peak of the static structure factor, and $\langle .\rangle$ denotes the ensemble average.  The relaxation time $\tau$ satisfies $F_s(q,\tau)=e^{-1}F_s(q,0)$.  By fitting the relaxation time measured at fixed temperature with the Vogel-Fulcher function, $\tau = \tau_0 {\rm exp} [M / (p_g - p)]$, where $\tau_0$ and $M$ are fitting parameters, we estimate the glass transition pressure $p_g$. The dynamical structure factors are measured from the time correlation of current \cite{shintani,hansen}:
\begin{equation}
S_{\lambda} (k, \omega) = \frac{k^2}{2\pi N \omega^2}\int_0^{\infty} dt \langle \vec{J}_{\lambda}(k, t) \cdot \vec{J}_{\lambda} (-k, 0) \rangle {\rm exp}({\rm i}\omega t),\label{s2}
\end{equation}
where
\begin{eqnarray}
\vec{J}_{L} (k, t) &=& \sum_j [(\vec{v}_j(t)\cdot \hat{k}) \hat{k}] {\rm exp} [{\rm i}\vec{k}\cdot \vec{r}_j(t)],\label{jl}\\
\vec{J}_{T} (k, t) &=& \sum_j [\vec{v}_j(t) - (\vec{v}_j(t)\cdot \hat{k}) \hat{k}] {\rm exp} [{\rm i}\vec{k}\cdot \vec{r}_j(t)],\label{jt}
\end{eqnarray}
where the sums are over all particles, and $\vec{v}_j$ is the velocity of particle $j$.  For simplifications, here we use the same symbol $S_{\lambda}(k,\omega)$ as that for $T=0$ \cite{note}.

%%%%%%%%%%%%%%%%%%%%%%%%%%%%%%%%%%%%%%%%%%%%%%%%%
\begin{figure}
\center
\includegraphics[width=0.48\textwidth]{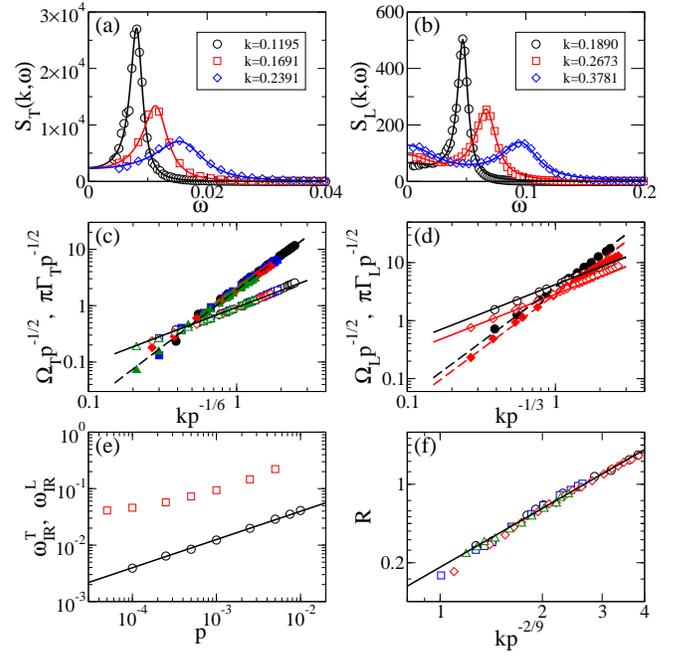}
\caption{\label{fig:fig1} (color online).  Analysis of the $T=0$ jammed solids.  (a) and (b) Examples of the dynamical structure factors $S_T(k,\omega)$ and $S_L(k,\omega)$ at $p=5\times 10^{-4}$ with the lines fittings with Eq.~(\ref{fit}). (c) and (d) Pressure scaled dispersion relation $\Omega_{\lambda}p^{-1/2}$ (empty symbols) and sound attenuation coefficient $\pi\Gamma_{\lambda}p^{-1/2}$ (solid symbols) against $kp^{-1/6}$ for the transverse ($\lambda=T$) and $kp^{-1/3}$ for the longitudinal ($\lambda=L$) excitations at $p=10^{-4}$ (circles), $5\times 10^{-4}$ (squares), $10^{-3}$ (diamonds), and $5\times 10^{-3}$ (triangles). The solid and dashed lines have a slope of $1$ and $2$, respectively. (e) Pressure dependence of the Ioffe-Regel frequencies $\omega_{IR}^T$ (circles) and $\omega_{IR}^L$ (squares). The solid line is the fitting with $\omega_{IR}^T\sim p^{1/2}$. (f) Ratio of the Rayleigh component to the Brillouin component, $R$, against $kp^{-2/9}$. The symbols are defined as in (c) and (d). The solid line has a slope of $3/2$.     }
\end{figure}
%%%%%%%%%%%%%%%%%%%%%%%%%%%%%%%%%%%%%%%%%%%%%%%%%%

Figures~\ref{fig:fig1}(a) and (b) show examples of the dynamical structure factors at $T=0$ calculated from Eqs.~(\ref{s1})-(\ref{ft}), which can be well fitted with \cite{hansen}
\begin{eqnarray}
S_{\lambda} (k,\omega) &=& S_{\lambda,B}(k,\omega) + S_{\lambda,R}(k,\omega) \label{fit} \\
&=&\frac{A_{\lambda}(k)}{[\omega^2 - \Omega_{\lambda}^2(k)]^2 + \omega^2 \Gamma_{\lambda}^2(k)} + \frac{B_{\lambda}(k)}{\omega^2+(D_{\lambda}k^2)^2}, \nonumber
\end{eqnarray}
where $S_{\lambda,B}(k,\omega)$ and $S_{\lambda,R}(k,\omega)$ are the Brillouin (solid) and Rayleigh (liquid) components, $A_{\lambda}(k)$ and $B_{\lambda}(k)$ are fitting parameters, $\Omega_{\lambda}(k)$ gives the dispersion relation, $\Gamma_{\lambda}(k)$ is the sound attenuation coefficient, and $D_{\lambda}$ is the thermal diffusivity.  For $S_T(k,\omega)$, the Brillouin component itself can fit the data well \cite{shintani,mizuno}, so $S_{T,R}(k,\omega)\approx 0$.  However, for $S_L(k,\omega)$, the Rayleigh component gives rise to the low-frequency increase with the decrease of the frequency, especially at large $k$ and low $p$.

The Rayleigh component of the longitudinal dynamical structure factor has been overlooked in most of previous studies of disordered solids.  Its presence implies that disordered solids still exhibit liquid characters at certain length scales.  Figure~\ref{fig:fig1}(f) shows the ratio of the Rayleigh contribution to the Brillouin contribution, $R(k)=\int_0^{\infty} d\omega S_{L,R}(k,\omega)/\int_0^{\infty} d\omega S_{L,B}(k,\omega)$, for the $T=0$ jammed solids at different pressures.  At each pressure, $R(k)\sim k^{3/2}$.  Interestingly, all curves collapse when we plot $R(k)$ against $kp^{-2/9}$, which implies a diverging length at point $J$ ($p=0$): $l^* \sim p^{-2/9}$.  This is consistent with the picture of jamming transition: The Rayleigh component dominates over infinitely large length scale, so that the jammed solids are on the verge of instability.

In Figs.~\ref{fig:fig1}(c) and (d), we illustrate how the Ioffe-Regel limit is determined.  As reported for some model glasses, $\Omega_{\lambda} \sim k$, and $\Gamma_{\lambda} \sim k^2$ \cite{shintani}.  The Ioffe-Regel limit is achieved when $\Omega_{\lambda}(k_{IR}^{\lambda})=\pi \Gamma_{\lambda}(k_{IR}^{\lambda})$, from which we obtain the Ioffe-Regel frequency $\omega_{IR}^{\lambda}=\Omega_{\lambda}(k_{IR}^{\lambda})$.  When $\omega>\omega_{IR}^{\lambda}$, a phonon undergoes multiple collisions before moving to a distance comparable to its wavelength and is thus ill-defined. At fixed pressure, $\omega_{IR}^L>\omega_{IR}^T$.

For the $T=0$ jammed solids, various characteristic frequencies have been defined, e.g., the onset frequency of the plateau in the density of vibrational states \cite{silbert2,wyart1,degiuli,xu2} or energy diffusivity \cite{xu3}.  It has been shown that these characteristic frequencies are scaled well with the pressure: $\omega^*\sim (\phi-\phi_{j0})^{(\alpha - 1)/2}\sim p^{1/2}$ \cite{silbert2,wyart1,xu2,xu3}, so at point $J$ $\omega^*$ decays to zero. It has also been argued that $\omega^*$ relates to the boson peak or Ioffe-Regel frequencies \cite{xu3,degiuli}.  However, no direct measure of the Ioffe-Regel frequencies has ever been performed for jammed solids.

In Fig.~\ref{fig:fig1}(e), we explicitly show the pressure dependence of the Ioffe-Regel frequencies for the $T=0$ jammed solids.  For the transverse one, $\omega_{IR}^T \sim p^{1/2}$, which vanishes at point $J$.  Because $\omega_{IR}^T=c_T k_{IR}^T$, where the transverse speed of sound $c_T= \sqrt{G/\rho}\sim p^{(\alpha - 3/2)/(2\alpha - 2)}$ \cite{ohern} with $G$ the shear modulus and $\rho$ the mass density, for systems with Hertzian repulsion, $\omega_{IR}^T\sim p^{1/3}k_{IR}^T$.  Therefore, there exists a diverging length approaching point $J$: $l_{IR}^T\sim (k_{IR}^T)^{-1} \sim p^{-1/6}$.  As verified in Fig.~\ref{fig:fig1}(c), $\Omega_T(k)$ and $\pi\Gamma_{T}(k)$ measured at different pressures collapse onto the same master curves with the same intersection when we plot $\Omega_T p^{-1/2}$ and $\pi\Gamma_T p^{-1/2}$ against $k p^{-1/6}$.

The longitudinal Ioffe-Regel frequency, however, shows the tendency to approach a nonzero value at point $J$, as shown in Fig.~\ref{fig:fig1}(e).  It has been argued that the vanishing of the characteristic frequency at point $J$ implies two possible diverging length scales, associated with the transverse and longitudinal excitations, respectively \cite{silbert2,wyart1,schoenholz}.  Here we show that $\omega_{IR}^L>0$ at point $J$, so the length $l_{IR}^L\sim (k_{IR}^L)^{-1}$ associated with the longitudinal excitations remains finite at point $J$.  The longitudinal speed of sound $c_L= \sqrt{(B+4G/3)/\rho}\sim \sqrt{B}\sim p^{(\alpha-2)/(2\alpha-2)}$, because the bulk modulus $B\gg G$ near point $J$.  If $\omega_{IR}^L\sim p^{1/2}$, we would observe scaling collapse of $\Omega_L(k)$ and $\Gamma_L(k)$ by plotting $\Omega_L p^{-1/2}$ and $\pi\Gamma_L p^{-1/2}$ against $k p^{-1/3}$, which is however not the case in Fig.~\ref{fig:fig1}(d).  Then, where will $\omega_{IR}^L$ vanish?

%%%%%%%%%%%%%%%%%%%%%%%%%%%%%%%%%%%%%%%%%%%%%%%%%
\begin{figure}
\center
\includegraphics[width=0.48\textwidth]{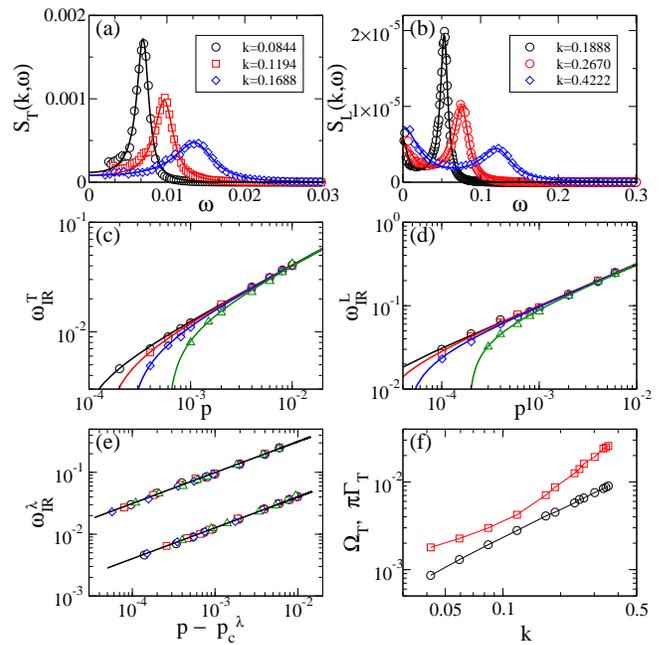}
\caption{\label{fig:fig2} (color online).  Analysis of glasses at $T>0$. (a) and (b) Examples of the dynamical structure factors $S_T(k,\omega)$ and $S_L(k,\omega)$ at $T=10^{-6}$ and $p=10^{-3}$ with the lines fittings with Eq.~(\ref{fit}). (c) and (d) Pressure dependence of the Ioffe-Regel frequencies $\omega_{IR}^T$ and $\omega_{IR}^L$ at $T=10^{-7}$ (circles), $5\times 10^{-7}$ (squares), $10^{-6}$ (diamonds), and $5\times 10^{-6}$ (triangles) with the lines fittings with $\omega_{IR}^{\lambda}\sim (p-p_c^{\lambda})^{1/2}$. (e) Scaling collapse of $\omega_{IR}^{\lambda}$ when plotted against $p-p_c^{\lambda}$ with the lines having a slope of $1/2$. The upper and lower branches are $\omega_{IR}^L$ and $\omega_{IR}^T$, respectively. (f) Dispersion relation $\Omega_T(k)$ (circles) and sound attenuation coefficient $\pi\Gamma_T(k)$ (squares) for a glass composed of $N=16384$ particles at $T=10^{-6}$ and $p=10^{-4}$ in $(p_g,p_j)$ (located by the star in Fig.~\ref{fig:fig3}). The lines are to guide the eye.
}
\end{figure}
%%%%%%%%%%%%%%%%%%%%%%%%%%%%%%%%%%%%%%%%%%%%%%%%%%

%%%%%%%%%%%%%%%%%%%%%%%%%%%%%%%%%%%%%%%%%%%%%%%%%
\begin{figure}
\center
\includegraphics[width=0.35\textwidth]{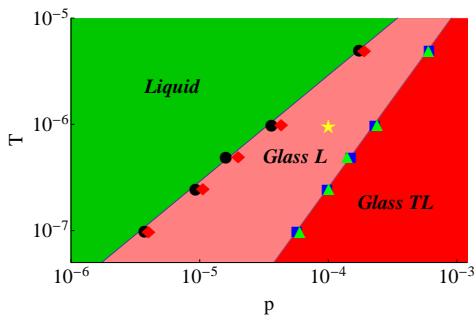}
\caption{\label{fig:fig3} (color online).  Phase diagram including the glass transition (circles with the line $T_g\sim p$) and jamming-like transition (squares with the line $T_j\sim p^{5/3}$). The diamonds and triangles locate the crossover temperatures $T_L$ ($\omega_{IR}^L=0$) and $T_T$ ($\omega_{IR}^T=0$), respectively. The star marks the location of the state shown in Fig.~\ref{fig:fig2}(f).
}
\end{figure}
%%%%%%%%%%%%%%%%%%%%%%%%%%%%%%%%%%%%%%%%%%%%%%%%%%

At $T=0$, mode analysis is inaccessible for unjammed systems below point $J$.  We can alternately measure excitations in low-temperature systems using Eqs.~(\ref{s2})-(\ref{jt}) and predict behaviors in the $T=0$ or hard sphere limit.  Figures~\ref{fig:fig2}(a) and (b) show examples of the dynamical structure factors measured at $T=10^{-6}$ and $p=10^{-3}$, which can also be well fitted with Eq.~(\ref{fit}).  We again estimate the Ioffe-Regel frequencies by locating the intersection between $\Omega_{\lambda}(k)$ and $\pi \Gamma_{\lambda}(k)$.

As shown in Figs.~\ref{fig:fig2}(c) and (d), at fixed pressure, both $\omega_{IR}^T$ and $\omega_{IR}^L$ decrease when increasing the temperature.  The decrease is faster at lower pressures, leading to the tendency that $\omega_{IR}^\lambda=0$ at a nonzero pressure $p_c^{\lambda}(T)$.  We find that all the curves can be well fitted with $\omega_{IR}^{\lambda}\sim (p-p_c^{\lambda})^{1/2}$, as illustrated by the scaling collapse of all the $T>0$ data in Fig.~\ref{fig:fig2}(e).  Does the vanishing of $\omega_{IR}^{\lambda}$ at $p_c^{\lambda}$ relate to any transitions?

As mentioned above, a thermal system with repulsion undergoes the glass transition and jamming-like transition in sequence under compression.  Figure~\ref{fig:fig3} is the $T-p$ phase diagram with both transitions.  As reported before, the glass transition temperature $T_g$ and jamming-like transition temperature $T_j$ are scaled well with the pressure: $T_g\sim p$ and $T_j\sim p^{\alpha/(\alpha - 1)}$ \cite{zhang,wang,xu1}.  For Hertzian repulsion, $T_j\sim p^{5/3}$.  In Fig.~\ref{fig:fig3}, we plot as well the temperatures $T_T$ and $T_L$ at which $\omega_{IR}^T$ and $\omega_{IR}^L$ tend to vanish, i.e., $p(T_{\lambda})=p_c^{\lambda}$.  Surprisingly, $T_T(p)$ collapses with $T_j(p)$, while $T_L(p)$ agrees well with $T_g(p)$.  This finding reveals the most significant underlying physics of the jamming-like transition: At fixed temperature, it is the lower limit for transverse phonons to survive. In the region $p_g(p_c^L)<p<p_j(p_c^T)$, only longitudinal phonons can be well-defined.  In the low-temperature limit this region is just the territory of hard-sphere glasses.

To further evidence that transverse phonons are ill defined in $(p_g,p_j)$, we show in Fig.~\ref{fig:fig2}(f) $\Omega_T(k)$ and $\pi\Gamma_T(k)$ of a state in the middle of this regime.  $\Gamma_T(k)$ is no longer scaled with $k^2$ at small $k$ and $\pi\Gamma_T(k)>\Omega_T(k)$ all the way down to $k=0$.  The jamming-like transition thus divides the whole glass regime into two parts.  In Fig.~\ref{fig:fig3}, we use ``Glass TL" and ``Glass L" to denote glasses with both transverse and longitudinal phonons at $p>p_j$ and with only longitudinal phonons at $p_g<p<p_j$.

\begin{figure}
\center
\includegraphics[width=0.45\textwidth]{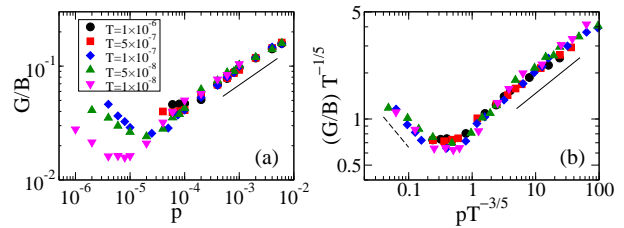}
\caption{\label{fig:fig4} (color online). (a) Pressure dependence of the ratio of the shear modulus to the bulk modulus, $G/B$, at various temperatures. (b) Scaling collapse of (a) by applying $G/B= T^{1/5} H(p/T^{3/5})$. The solid lines have a slope of $1/3$, while the dashed line has a slope of $-1/2$.}
\end{figure}

In the solid perspective, the absence of well-defined transverse phonons should be the intrinsic cause of the distinct material properties of Glass L from Glass TL \cite{wang}.  In Fig.~\ref{fig:fig4}(a) we show the ratio of the shear modulus to the bulk modulus, $G/B=(c_L^2/c_T^2-4/3)^{-1}$, at various temperatures and pressures.  In the $T=0$ jammed solids, it is known that $G/B\sim(\phi-\phi_{j0})^{1/2}\sim p^{1/(2\alpha -2)}$ \cite{ohern,basu,tighe,goodrich}.  For Hertzian repulsion, $G/B\sim p^{1/3}$.  Figure~\ref{fig:fig4}(a) shows that $G/B\sim p^{1/3}$ when $p>p_j$, while such a scaling relation fails at $p<p_j$.  Interestingly, all the curves in Fig.~\ref{fig:fig4}(a) collapse well onto the same master curve with two distinct regimes when the scaling function $G/B= T^{1/5} H(p/T^{3/5})$ is applied, as shown in Fig.~\ref{fig:fig4}(b).  The scaling function is constructed from the known scaling $p_j\sim T^{3/5}$ \cite{zhang,wang,berthier3}.

%%%%%%%%%%%%%%%%%%%%%%%%%%%%%%%%%%%%%%%%%%%%%%%%%
%%%%%%%%%%%%%%%%%%%%%%%%%%%%%%%%%%%%%%%%%%%%%%%%%%

In contrast to $G/B\sim p^{1/3}$ at $p>p_j$, $G/B\sim (p/T)^{-1/2}$ at $p<p_j$ and low temperatures with a short transition region near $p_j$.  The opposite pressure dependence of $G/B$ on both sides of $p_j$ directly reflects the unusual vibrational properties of Glass L.  In the hard sphere limit, $G/B$ of hard sphere glasses decreases with increasing the packing fraction or pressure and decays to zero at point $J$, where $p/T\rightarrow \infty$, consistent with the divergence of the mean squared displacement approaching point $J$ \cite{berthier3}.

Our major findings that Glass L can only support longitudinal phonons and its distinct pressure dependence of $G/B$ from Glass TL appeal for the reconsideration of the nature of hard sphere glasses.  They also validate the jamming-like transition, which converges to point $J$ in the $T=0$ limit, to be a physically meaningful transition and a possible substitute of the glass transition as the boundary of the jamming phase diagram in the $T-\phi$ plane \cite{liu}: It is the transition to ``normal" disordered solids with both transverse and longitudinal phonons.  

The conclusions drawn here are valid for low-temperature systems with repulsions near point $J$.  They may not be simply generalized to systems with long-range attractions, because the jamming transition is inaccessible to these systems.  Furthermore, recent studies have demonstrated that repulsive and attractive systems differ in their glassy dynamics and vibrational properties at low densities \cite{berthier4,zhang1,wang1}, which magnify when lowering the density and are thus the direct manifestations of their distinct awareness of the existence of point $J$.

We are grateful to Ludovic Berthier for the critical reading.  This work is supported by National Natural Science Foundation of China Grant No. 21325418, National Basic Research Program of China (973 Program) Grant No. 2012CB821500, CAS 100-Talent Program Grant No. 2030020004, and Fundamental Research Funds for the Central Universities Grant No. 2340000034.


\begin{thebibliography}{50}
\bibitem{note} Note that $S_{\lambda}(k,\omega)$ in Eq.~(\ref{s2}) includes the contribution of the thermal energy and is proportional to $k_BT$, while it in Eq.~(\ref{s1}) is scaled by $k_BT$.  This difference only affects the magnitude of $S_{\lambda}(k,\omega)$. It does not affect the measure of the dispersion relation and sound attenuation coefficient.
\end{thebibliography}
\end{document}